\documentclass[12pt]{article}
\begin{document}
\thispagestyle{empty}
\begin{flushright}
{\tt OSU-HEP-01-01}\\
{\tt hep-ph/0101046 }\\
January 2001 \\
\end{flushright}
\vspace*{2cm}
\begin{center}
{\baselineskip 25pt
\large{\bf 
Solving the $\mu$ Problem in Gauge Mediated \\[0.1in]
Supersymmetry Breaking Models with
Flavor Symmetry
}}

\vspace{1cm}

{\large  K.S. Babu}\footnote
{email: {\tt babu@osuunx.uuc.okstate.edu}}
and {\large  Yukihiro Mimura}\footnote
{email: {\tt mimura@okstate.edu}}

\vspace{.5cm}

{{\it Physics Department, Oklahoma State University,
             Stillwater, OK 74078}}

\vspace{.5cm}

\vspace{1.5cm}

{\bf Abstract}

\end{center}
\vspace{0.5cm}

We suggest a solution to the $\mu$ problem of gauge mediated supersymmtery
breaking models based on flavor symmetries.  In this scenario the $\mu$ term 
arises through the vacuum expectation value of a singlet scalar 
field which is suppressed by a flavor symmetry factor relative to the scale 
of dynamical SUSY breaking.  The same flavor symmetry also ensures that the
soft SUSY breaking parameter $B\mu$ is not much larger than $\mu^2$, a necessary
condition for the stability of electroweak symmetry breaking.  Explicit 
examples where $B\mu \sim \mu^2$ and $B\mu \ll \mu^2$ are presented.  The
latter case provides a natural solution to the supersymmetric CP problem.  
We show that the same flavor symmetry that suppresses the $\mu$ and the
$B\mu$ parameters can also play a role in explaining the fermion mass and 
mixing hierarchy puzzle.

\bigskip
\newpage

\baselineskip 20pt

\noindent {\large \bf \S 1}

There is a huge hierarchy between the electroweak scale ($\sim 10^2$ GeV)
and the Planck scale ($\sim 10^{19}$ GeV). Supersymmetric theories are 
excellent candidates which can stabilize this hierarchy against
quantum corrections.  
Furthermore, these theories can also explain the origin of 
such a huge hierarchy
in  scenarios where the electroweak symmetry is broken
radiatively \cite{Inoue}.

However, supersymmetry (SUSY) alone is not sufficient to explain electroweak 
symmetry breaking.  For it to be successful, 
the supersymmetry breaking mass parameters should  lie
around the electroweak breaking scale.  This is needed for a natural
resolution of the hierarchy problem.  In the minimal supersymmetric standard model
(MSSM) with radiative electroweak symmetry breaking, the stability of the
Higgs potential requires the SUSY breaking masses as well as the 
supersymmtric Higgs mass parameter $\mu$  to be around the electroweak scale.  
Here the superpotential term $W \ni 
\mu H_u H_d$, where $H_u$ and $H_d$ are the two Higgs doublets of MSSM,
defines the $\mu$ parameter.  The soft SUSY breaking $B\mu$ parameter,
defined through $V \ni B\mu H_u H_d $, should also lie around the
electroweak scale.  These requirements are evident from the two minimization
conditions of the MSSM Higgs potential, which read at tree--level as:
\begin{equation}
M_Z^2 = - \mu^2 + \frac{m_{H_d}^2 - m_{H_u}^2 \tan^2 \beta}{\tan^2 \beta -1}~,
\end{equation}
\begin{equation}
\sin 2\beta = \frac{2 B \mu}{2 \mu^2 + m_{H_u}^2 + m_{H_d}^2}~.
\end{equation}
Here we have used standard notation with $\tan\beta = v_u/v_d$, $v_u$ and 
$v_d$ being the vacuum expectation values (VEVs) of the Higgs fields
$H_u$ and $H_d$.  Eq. (1) clearly shows that $\mu^2$ cannot be much
larger than $M_Z^2$ without fine--tuning, Eq. (2) shows that $B\mu$ 
is bounded by $\mu^2 + (m_{H_u}^2+m_{H_d}^2)/2$.  

Since the $\mu$ parameter is part of the supersymmetric Lagrangian,
a question arises as to why it is not much above the electroweak scale,
say near the Planck scale.  Such a large value of $\mu$ is of 
course inconsistent with symmetry breaking requirements.    
This is the so--called $\mu$ problem.  
There is a good solution \cite{Giudice} for this problem in scenarios where
supersymmetry
breaking is communicated to the standard model sector through 
gravity \cite{Barbieri}.  In supergravity models, 
the supersymmetry breaking masses are given by
$m_{SUSY} \sim F_X/M_P$, where $F_X$, the order parameter for
SUSY breaking, is the $F$ component of a
chiral spurion field $X$.  Since the $\mu$ term is not
invariant under a $U(1)_{PQ}$ symmetry, it is conceivable that it
is zero to begin with.  
If the K\"ahler potential contains the terms
\begin{equation}
K = (\frac{X^\dagger}{M_P} + \frac{X X^\dagger}{M_P^2} + \cdots)~ H_u H_d~,
\end{equation}
then a $\mu$ term of the correct magnitude
is generated as $\mu \sim F_X/M_P \sim m_{SUSY}$.  At the
same time, $B \mu \sim (F_X/M_P)^2 \sim m_{SUSY}^2$ is induced.
Thus $\mu^2$ and $B \mu$ are naturally of the same order in this scenario,
consistent with electroweak symmetry breaking requirements.  

While supergravity models can provide an elegant solution to the $\mu$
problem, it is not at all clear if they can suppress flavor violation mediated
by the exchange of supersymmetric particles to adequate levels.
Gauge mediated supersymmetry
breaking models (GMSB) \cite{gmsb} have a great advantage in this regard, 
since they are manifestly flavor--conserving.  
In GMSB models, the supersymmetry breaking masses are generated as 
$m_{SUSY} \sim (\alpha/4\pi)F_X/\left\langle X \right\rangle$ \cite{gmsb,dnns,gr} where 
$\left\langle X\right\rangle$, the scalar component
of a chiral superfield $X$, is much smaller than $M_P$.
In this scenario, the mechanism of supergravity models
for generating $\mu$ and $B \mu$ parameters does not work well 
since their magnitudes are too small.  The same mechanism that
suppresses flavor violation also suppresses the magnitude of $\mu$
arising through Eq. (3).

Any attempt to solve the $\mu$ problem 
should circumvent the following challenges. Suppose that
the $\mu$ term arises through the superpotential coupling $W \ni \lambda
H_u H_d X$, where $X$ is the spurion field whose $F$--component
breaks supersymmetry. 
Then $\mu = \lambda \left\langle X \right\rangle$.
For the SUSY breaking masses of the standard model
sector to be of order $10^2-10^3$ GeV, it is necessary that
$F_X/\left\langle X\right\rangle \sim 10^4 - 10^5$ GeV.  The
same superpotential coupling that induces a $\mu$ term
will also induce a $B\mu$ term given by 
$B\mu = \lambda F_X$,
so that $B = F_X/\left\langle X \right\rangle \sim 10^4-10^5$ GeV \cite{pomarol}. 
Such a value of $B$ is inconsistent with symmetry breaking constraint, Eq. (2).

If a separate singlet scalar which does not couple
to the dynamical SUSY breaking sector (so that its $F$--component
is small)  is employed to induce the $\mu$ term
through its superpotential coupling
to $H_uH_d$ (as in the next to minimal supersymmetric model \cite{murayama}),
the resulting model will be inconsistent with phenomenology, 
since such a scenario will lead to
light scalars excluded by experimental data.  For a variation of this
model, see Ref. \cite{han}. 

There have been several attempts to solve the $\mu$ problem in the 
context of GMSB models \cite{mu}.  Some of the ideas 
discussed involve use of non--renormalizable
terms, gravity contribution, contribution from an extra $U(1)$ 
gauge symmetry or a separate dynamical
sector to generate the $\mu$ term.  

In this paper we propose a solution to the $\mu$ problem which does not face the
above--mentioned difficulties.  We use flavor symmetries 
to suppress the magnitude of $\mu$ relative to the dynamical SUSY breaking 
scale.  These symmetries also ensure that $B \mu \leq \mu^2$, so that 
electroweak symmetry breaking is not destabilized.  We present an explicit example which
uses additional singlet superfields which acquire flavor symmetry--suppressed
VEVs for their scalar and $F$ components, and use these fields to induce
the $\mu$ parameter.  The same flavor 
symmetries that suppress $\mu$ and $B$ parameters 
can be used to explain the pattern of quark and lepton masses 
and mixings, as we will show.

\vspace{0.5cm}
\baselineskip 20pt
\noindent{\large \bf \S 2} 

To see how the proposed mechanism works, 
we build on the minimal messenger $U(1)$ model of
Ref. \cite{dnns}.  Consider first the minimal
version of the model as discussed in \cite{dnns}.  
The superpotential of the messenger sector is 
\begin{equation}
W = k_1 X \varphi_+ \varphi_- - {k_2 \over 3} X^3 + k_3 X M \bar M~,
\end{equation}
where $X$ is a singlet neutral under the messenger $U(1)_m$ and
$\varphi_\pm$ have $U(1)_m$ charges $\pm 1$.
$M$, $\bar M$ are messenger fields
carrying standard model quantum numbers (eg: ${\bf 5}+\bar{\bf 5}$
of $SU(5)$).
The $\varphi_\pm$ fields couple to the dynamical SUSY breaking sector
and through loops receive a SUSY breaking negative mass--squared 
$V \ni -m_\varphi^2(|\varphi_+|^2+|\varphi_-|^2)$.  The scalar potential 
also contains a $U(1)_m$~ $D$--term given as $V \ni {g_m^2 \over 2}
(|\varphi_+|^2-|\varphi_-|^2)^2$.  An $R$--symmetry prevents possible
bare mass terms in Eq. (4).  

The couplings $k_{1,2,3}$ of Eq. (4) can be made real and positive
by field redefinitions.  
There is then a CP--conserving local minimum, with $\left\langle M \right
\rangle = \left \langle \bar{M} \right\rangle = 0, ~\left\langle |\varphi_+|^2
\right\rangle = \left\langle |\varphi_-|^2\right\rangle \equiv \varphi^2$,
and the VEVs given by

\begin{equation}
\left\langle X^2 \right \rangle
 = \frac{k_2 - k_1}{k_1^2 (2k_2 - k_1)} m_\varphi^2~,
\quad 
\varphi^2 = \frac{k_2^2}{k_1^3 (2 k_2 - k_1)} m_\varphi^2~.
\quad
\label{local_minimal}
\end{equation}

\noindent
The stability of this local minimum is ensured if the following conditions
are satisfied \cite{randall}\footnote{The potential has a deeper minimum which preserves
supersymmetry
where $M,\bar{M}$ have non--vanishing VEVs.  For a discussion of cosmological
tunneling rate from the desirable local minimum of Eq. (5) into this 
true minimum, see Ref. \cite{randall}.}:

\begin{equation}
k_2 > k_1,~~~ 2 g_m^2 >  {k_1^3 \over k_2},~~~ k_3^2 > {k_1 k_2 \over k_2-k_1}~.
\end{equation}
The $F$--component of $X$ is
\begin{equation}
\left\langle F_X \right\rangle = 
\frac{k_1 k_2}{k_1^2 (2k_2 - k_1)} m_\varphi^2~.
\end{equation}

The mass spectrum of the messenger fields
$M,\bar{M}$ is SUSY--noninvariant, owing to
$\left \langle F_X \right\rangle \neq 0$.
These fields convey SUSY breaking 
to the MSSM chiral superfields and the gauginos
through strong and electroweak loops.  The magnitude of these
SUSY breaking masses is governed by the parameter $\Lambda \equiv 
\left\langle F_X \right\rangle/\left\langle X \right\rangle$ which
must lie in the range $\Lambda \sim (10^4-10^5)$ GeV for squark and
gaugino masses to be in the range of $(10^2-10^3)$ GeV.
The model thus provides a natural mechanism for mediating SUSY
breaking, while also preserving flavor conservation in the SUSY
breaking sector.  The phenomenology of these minimal models have been
discussed extensively in the literature \cite{pheno,bkw,wells,dutta}.

The model does not however address the origin of
the $\mu$ parameter.  
As already noted, if the $\mu$ term is induced through 
the superpotential coupling $\lambda H_u H_d X$, 
a $B$ term given by $B = \left\langle F_X \right\rangle/\left\langle X 
\right\rangle
= \Lambda \simeq (10^4-10^5)$ GeV will result, which is inconsistent with
the electroweak symmetry breaking condition, Eq. (2).

Consider now an extension of the minimal model
with a $U(1)_F$ flavor symmetry, which is broken near the Planck scale by
a pair of chiral superfields $\Phi$ and 
$\bar \Phi$ carrying $U(1)_F$ charges of $+1$ and $-1$ respectively.
The VEV ratios $\left\langle \Phi \right \rangle/M_*, \left\langle \bar{\Phi} 
\right\rangle/M_* \ll 1$, where $M_*$ is the reduced Planck scale,
provide the small parameter which will explain
the smallness of $\mu$ and $B\mu$.  In addition to the $X$ field, 
we introduce two singlets $Y,Z$.  The 
$U(1)_F$ charges of $X$, $Y$, $Z$ are $0,~ 2,~ -4$ respectively.

The superpotential of the model now has two sets of terms: One piece given by
$W$ of Eq. (4), and another piece given by
\begin{eqnarray}
\Delta W &=&  a_1 Y^2 Z - a_2 \lambda X^2 Y + a_3 \lambda^2 X Y^2 - 
a_4 \lambda^2 X^2 Z +
a_5 \lambda \varphi_+\varphi_- Y + \nonumber \\ &~& a_6 \lambda^2 
\varphi_+\varphi_-Z 
         + a_7 \lambda^3 Y^3 + a_8 \lambda^3 Y Z^2 + a_9 \lambda^4 X Z^2 + a_{10} 
\lambda^6 Z^3~. 
\end{eqnarray}
Here $\lambda \equiv (\left\langle \Phi\right\rangle/M_*)^2 \ll 1$ is the
flavor symmetry breaking parameter. The couplings $k_i$ and $a_i$ of Eqs. (4),
(8) are assumed to take natural values of order one.  The $Y$ and $Z$ fields
also couple to th messenger sector through $\lambda Y M \bar{M} + \lambda^2 Z
M \bar{M}$.  These superpotential couplings will not however play any role 
in our analysis of the local minimum.  

Although the parameters of Eq. (4) and (8) cannot be all made real, we will
analyze the potential assuming that all of $a_i$ and $k_i$ are real.  This
will enable us to look for a CP conserving local minimum.  

The scalar potential of the model is given by
\begin{eqnarray}
V &=& |F_X|^2 + |F_Y|^2 + |F_Z|^2 + |k_1 X + a_5 \lambda Y + a_6 \lambda^2 Z|^2
 (|\varphi_+|^2 + |\varphi_-|^2) \nonumber \\
&+& {g_m^2 \over 2} (|\varphi_+|^2 - |\varphi_-|^2)^2 - m^2_\varphi 
(|\varphi_+|^2 + |\varphi_-|^2)
+ |k_3 X|^2 (|M|^2 + |\bar M|^2)  \nonumber \\
&+& {\rm D~terms~for~}M,\bar{M}~{\rm fields} ~.
\end{eqnarray}

\noindent Here $F_X$, $F_Y$, and $F_Z$ are given by
\begin{eqnarray}
F_X &=& k_1 \varphi_+ \varphi_- - k_2 X^2  + k_3 M \bar{M} -2 a_2 \lambda X Y + a_3 \lambda^2 Y^2
      - 2 a_4 \lambda^2 X Z + \cdots \\
F_Y &=& 2 a_1 Y Z - a_2 \lambda X^2 + 2 a_3 \lambda^2 X Y + a_5 \lambda
\varphi_+\varphi_- + \cdots \\
F_Z &=& a_1 Y^2 - a_4 \lambda^2 X^2  + a_6 \lambda^2 \varphi_+\varphi_- + \cdots
\end{eqnarray}
where the dots represent terms of order $\lambda^3$ and higher ($a_7-a_{10}$
terms of Eq. (8)), which turn out
to be insignificant for our analysis.  

In the $\lambda \rightarrow 0$ limit, the potential has a local minimum
which is the same as
Eq. (\ref{local_minimal}) along with $Y=0$ and arbitrary $Z$.  We seek a minimum
which is a small perturbation around this one.  We will see that including
terms of order $\lambda$ and $\lambda^2$, this minimum is shifted so that
$\left\langle Y \right\rangle \sim \lambda \left \langle X\right\rangle$ 
and $\left\langle Z\right \rangle \sim \left \langle X \right\rangle$.
At this minimum, $\left\langle F_Y \right\rangle \sim \lambda \left\langle
X^2 \right\rangle$ and $\left\langle F_Z\right\rangle \sim \lambda^2 
\left\langle X^2 \right\rangle$, as we will see.  Such a minimum is what
we need to explain the smallness of $\mu$ and $B\mu$ parameters.  
As before, we will take $\left\langle M \right\rangle = \left\langle \bar{M}
\right\rangle = 0$.  

Since $\lambda$ is a small perturbation to Eq. (4), the minimization conditions
of Eq. (5) and the stability conditions of Eq. (6) are essentially unaffected
by the inclusion of $\Delta W$ (Eq. (8)).  Minimization with respect to $Y$ and
$Z$ lead to the conditions
\begin{eqnarray}
 \frac14 \frac{\partial V}{\partial Y} &\simeq& -a_2 \lambda X F_X^* + a_1 Z F_Y^*+
{a_5 \over 2} \lambda (k_1X)^*(|\varphi_+|^2+|\varphi_-|^2) = 0 \\
\frac14 \frac{\partial V}{\partial Z} &\simeq& -a_4 \lambda^2 X F_X^* + a_1 Y F_Y^* 
+{a_6 \over 2} \lambda^2 (k_1X)^*(|\varphi_+|^2+|\varphi_-|^2) = 0~.
\end{eqnarray}
Using Eqs. (5) and (7), these lead to 
\begin{equation}
\left\langle Y \right\rangle \simeq \lambda \left\langle Z
\right\rangle \left({a_4k_1-a_6k_2 \over a_2k_1-a_5k_2}\right)~, 
\end{equation}
and a cubic equation in $\left\langle Z\right\rangle/\left\langle X\right\rangle$:
\begin{equation}
\left[{\left\langle Z \right\rangle \over \left\langle X \right\rangle}\right]
^3 - A {\left\langle Z \right\rangle \over \left\langle X \right\rangle}
-{1 \over 2 a_1^2}{k_2 \over k_2-k_1}{(a_2k_1-a_5k_2)^2 \over (a_4k_1-a_6k_2)}
= 0 ~,
 \end{equation}
where the parameter $A$ is defined as
\begin{equation}
A \equiv {a_2 \over 2 a_1}\left({a_2 k_1 - a_5 k_2 \over a_4 k_1-a_6k_2}\right)\left[
1 - {a_5 k_2^2 \over a_2k_1(k_2-k_1)}\right]~.
\end{equation}

The second derivatives of $V$ with respect to $Y$ and $Z$ at the minimum are
\begin{eqnarray}
\frac14 \frac{\partial^2 V}{\partial Y^2} &=&
2 a_1^2 \left\langle Z\right\rangle^2~, \\
\frac14 \frac{\partial^2 V}{\partial Z^2} &=&
2 a_1^2 \left\langle Y\right\rangle^2~, \\
\frac14 \frac{\partial^2 V}{\partial Y \partial Z} &=&
4 a_1^2 \left\langle Y\right\rangle 
\left\langle Z\right\rangle - a_1 a_2 \lambda \left\langle 
X\right\rangle^2 + a_1 a_5 \lambda \varphi^2~.
\end{eqnarray}
For the stability of the local minimum it is necessary that 
$\frac{\partial^2 V}{\partial Y^2} \frac{\partial^2 V}{\partial Z^2}
 - |\frac{\partial^2 V}{\partial Y \partial Z}|^2 > 0$.
Thus we need
\begin{equation}
\left[6 a_1 \left\langle Y\right\rangle \left\langle Z\right\rangle - 
a_2 \lambda \left\langle X\right\rangle^2+a_5 \lambda \varphi^2\right] 
\left[-2 a_1 \left\langle Y\right\rangle
\left\langle Z\right\rangle + a_2 \lambda \left\langle X\right\rangle^2
-a_5\lambda \varphi^2\right] > 0~.
\end{equation}

\noindent 
This constraint implies the following restriction on $\left\langle Z\right
\rangle/\left\langle X\right\rangle$:
\begin{equation}
{A \over 3} \leq \left[{\left\langle Z \right\rangle \over \left\langle X \right\rangle}
\right]^2 \leq A
\end{equation}
with $A$ as defined in Eq. (17).  
A necessary condition to satisfy Eq. (22)  simultaneously with Eq. (16) is
\begin{equation}
A^{3/2} \ge {3 \sqrt{3} \over 4 a_1^2}{k_2 \over k_2-k1}{(a_2k_1-a_5k_2)^2 \over
|a_4k_1-a_6k_2|}~.
\end{equation}

As it turns out, it is sufficient to examine the stability of the potential
in the $(Y,Z)$ subspace.  The mixing of $(Y,Z)$ with $(X,\varphi_+,\varphi_-)$
are suppressed by powers of $\lambda$ and does not affect the eigenvalues
of the second derivative matrix to leading order.  For example, the mixings of
$Y$ with $(X, \varphi_+,\varphi_-)$ are all of order $\lambda$, while the
mass--squared of $Y$ field is of order one.  

It is not difficult to find range of parameters of the model where the desired
local minimum is a stable configuration.  Consider the choice of parameters
where all couplings are equal to $+1$, except for $k_1$ and $a_1$, which
are respectively $0.3$ and $-1$.  In this case, $A = 1.88$ and Eq. (16)
solves for $\left\langle Z\right\rangle/\left\langle X\right\rangle = 1.21$.
This value is consistent with the constraint of Eq. (22).  (The right--hand side
of Eq. (23) is 1.30 in this case, which is less than $A^{3/2}$.)
We note that the
mass of the singlet $Z$ is $\sim \lambda\left\langle X \right\rangle$ which
is somewhat suppressed, but may still be in the several TeV range.  

\vspace{0.5cm}

\noindent {\large \bf \S 3} 

The stability of the local minimum being established, 
let us now turn to the origin of $\mu$ and $B\mu$ parameters in the
model just described.  

Defining the $U(1)_F$ charge of the combination $H_u H_d$ to be $2c$,
we can write the superpotential terms that will induce the $\mu$ parameter as
\begin{equation}
W = (\lambda^{|c|} X + \lambda^{|1+c|} Y + \lambda^{|c-2|} Z)~ H_u H_d~.
\end{equation}
Here the first term for example,  
arises either from $H_uH_dX\Phi^n$ or from $H_uH_dX\bar{\Phi}^n$
with $n$ an integer. The other two terms have a similar origin.
Note that in all cases $2c$ must be an integer.   
Recall that at the local minimum of the potential we have 
\begin{equation}
\left\langle Y\right\rangle \sim \lambda \left\langle X\right\rangle, 
\quad \left\langle F_Y\right\rangle \sim \lambda \left\langle X
\right\rangle^2, \quad \left\langle Z\right\rangle \sim \left\langle X
\right\rangle, \quad \left\langle F_Z\right\rangle \sim \lambda^2 
\left\langle X\right\rangle^2~.
\end{equation}

\noindent
Thus the $\mu$ parameter is
\begin{equation}
\mu \sim (\lambda^{|c|} + \lambda^{1+|1+c|} + \lambda^{|2-c|}) \left\langle 
X\right\rangle~,
\end{equation}
and the $B \mu$ parameter is
\begin{equation}
B \mu \sim (\lambda^{|c|} + \lambda^{1+|1+c|} + \lambda^{2+|2-c|}) 
\left\langle X\right\rangle^2~.
\end{equation}
Here again we note that $\lambda$ is $(\Phi/M_*)^2$ or $(\bar \Phi/M_*)^2$.

From Eqs. (26) and (27) it is clear that $c < 0$ will lead to $B\mu \gg
\mu^2$, which is inconsistent with electroweak symmetry breaking condition.
So we focus on the case $c > 0$. If $c < 1$, again $B\mu \gg \mu^2$.  
Thus we are left with two possibilities.
\begin{itemize}
\item[(i)] $1< c < 2$: In this case we have
\begin{equation}
\mu \sim \lambda^{2-c} X, \qquad B\mu \sim \lambda^c X^2,
\end{equation}
so that
\begin{equation}
\frac{B\mu}{\mu^2} \sim \lambda^{3c-4}.
\end{equation}
Thus we need $c \geq \frac43$ in this case.

\item[(ii)] $c>2$: Here we have
\begin{equation}
\mu \sim \lambda^{c-2} X, \qquad B\mu \sim \lambda^c X^2,
\end{equation}
and
\begin{equation}
\frac{B\mu}{\mu^2} \sim \lambda^{4-c}.
\end{equation}
Thus in this case $c$ must be smaller than $4$.

\end{itemize}

To summarize, we have a class of successful models parametrized by $c$ in the range
\begin{equation}
\frac43 \leq c \leq 4 \qquad (c \neq 2)~,
\end{equation}
where both $\mu$ and $B\mu$ are in the desired range.  
The case of $c=2$ is not favored since in this case $\mu$ is not
suppressed relative to $\left\langle X \right\rangle$ (Cf. Eq. (26)).  Combining
Eq. (32) with the restriction that $2c$ must be an integer, we arrive at
\begin{equation}
c = ({3 \over 2}, ~{5 \over 2},~ 3,~ {7 \over 2},~ 4)~.
\end{equation}

In the case where $c=4$, we find $B\mu/\mu^2 = {\cal O}(1)$.
This case will lead to the minimal messenger models with unconstrained
value of the parameter $\tan\beta$.  On the other hand, if $c \neq 4$,
we have $B\mu \ll \mu^2$.  This class of models \cite{bkw} is attractive since it
will naturally lead to
large value of $\tan\beta$ (thus explaining the large top--bottom mass
hierarchy) and also provide a simple resolution for the
SUSY CP problem \cite{dns}.   Since gauge mediated SUSY breaking models have generically
suppressed values for the trilinear $A$ terms, the smallness of $B$ would imply
that the electric dipole moments of the neutron and the electron will be small,
suppressed by a factor $(B\mu/\mu^2)$.  Take for example the choice $c=3$.
In this case $\mu \sim \lambda \left\langle X \right\rangle$ and
$B\mu \sim \lambda^3 \left\langle X \right\rangle^2$ so that $B\mu/\mu^2
\sim \lambda$.  If $\left\langle X \right\rangle \sim 30$ TeV and $\mu
\sim 300$ GeV, then $\lambda \sim 10^{-2}$, which will be the suppression
factor for the electric dipole moments.  

The decay $b \rightarrow s \gamma$ can receive significant 
corrections if $B\mu/\mu^2 \ll 1$ \cite{wells,sarid}, as $\tan\beta$
is large in this case.    
If the messenger mass scale is low, $\sim (10^5-10^7)$ GeV,
then the SUSY contribution to $b \rightarrow
s \gamma$ interferes destructively with the standard model contribution.
The resulting constraints have been analyzed and shown to
be consistent with present data in Ref. \cite{sarid}.

It is possible to identify the $U(1)_F$ symmetry with flavor
symmetries usually associated with explaining quark and lepton mass
hierarchy.  Consider the case where  
$\left\langle X\right\rangle \sim 10^5$ GeV and $\mu \sim10^2$ GeV.
We need $\lambda^{|2-c|} \sim 10^{-3}$ then.
Take $c=4$ as an example.  In this case $\lambda^2$ is about $10^{-3}$ and 
$\left\langle\Phi\right\rangle/M_*$ is about $1/6$.
This is appropriate for building quark/lepton mass hierarchy by using 
the $U(1)_F$ symmetry \cite{flavor}.  Since the $U(1)_F$ charges of the
standard model fermions are not determined in our analysis, there is a
lot of flexibility in constructing realistic fermion mass hierarchy models.  

There is a simple generalization of the model presented here
which allows for even more flexibility in applying it
to the fermion mass hierarchy.  We observe that the superpotential
of Eq. (4) and (8) is invariant if the $(X,Y,Z)$ superfields have
$U(1)_F$ charges of $(0,n,-2n)$ where $n$ is any positive integer.  
Then Eq. (8) will remain unchanged, provided that the definition of $\lambda$ 
is modified to $\lambda \equiv (\left\langle \Phi \right\rangle/M_*)^n$.
The discussions of the local minimum of Sec. 2 will not be altered.  In the
discussion of $\mu$ and $B\mu$ parameters of this section, $c$ will get replaced by
$2c/n$ with $2c$ still defined as the integer $U(1)_F$ charge of $H_uH_d$.  
Thus Eq. (32) will be replaced by $4/3 \leq 2c/n \leq 4 ~(2c/n \neq 2)$.  
In the example discussed in the previous paragraph, 
$\lambda^2 \sim 10^{-3}$ will correspond to $\left \langle \Phi \right\rangle/M_*
\sim 10^{-3/(2n)}$.  Numerically $\left\langle \Phi \right\rangle/M_* \sim
(1/36,~1/6,~ 1/3.3)$ for $n=(1,~2,~3)$.

\vspace{0.5cm}

\noindent{\large\bf \S 4}

In summary, we have suggested in this paper a solution to the $\mu$
problem in the context of gauge mediated supersymmetry breaking models.
It makes use of flavor symmetries to suppress $\mu$ and the soft SUSY
breaking parameter $B\mu$ relative to the scale of dynamical SUSY breaking.
We have analyzed the scalar potential in a concrete model and have found
that stable local minima are possible where some of the fields acquire
VEVs and $F$--components which are suppressed relative to the
dynamical SUSY breaking scale by a flavor symmetry factor.
The flavor symmetry also helps us to 
circumvent the generic problem of having $B\mu \gg \mu^2$
in GMSB models.  Since these fields with suppressed VEVs also have 
couplings to the messenger sector, there are no unwanted light scalars
in the theory.  We have found explicit realizations of models where
the SUSY breaking parameters $A \approx 0$, $B \approx 0$, which solves
the SUSY CP problem in a simple way.  The flavor symmetry that suppresses
$\mu$ and $B\mu$ parameters can also be used to explain the fermion mass
hierarchy puzzle.

\section*{Acknowledgments}

We wish to thank B. Dutta, S. Nandi and J. Wells for useful discussions.
This work is supported in part by DOE Grant \# DE-FG03-98ER-41076.  The
work of K.B is supported also in part by a grant from the Research Corporation,
DOE Grant \# DE-FG02-01ER4864 and by the OSU Environmental Institute.

\end{document}